\documentclass[10pt,showpacs,amsmath,amssymb]{article}
\usepackage{amsfonts}
\usepackage{mathdots}
\usepackage{color}

\newcommand{\field}[1]{\mathbb{#1}}

\begin{document}


\title{Constant curvature solutions of Grassmannian sigma models: (2) Non-holomorphic solutions}

\author{L. Delisle${}^{1,4}$, V. Hussin${}^{1,2,5}$ and W. J. Zakrzewski${}^{3,6}$}

\footnotetext[1]{D\'epartement de Math\'ematiques et de
Statistique, Universit\'e de Montr\'eal, C.P. 6128,
Succ.~Centre-ville, Montr\'eal (Qu\'ebec) H3C 3J7, Canada.}
\footnotetext[2]{Centre de Recherches Math\'ematiques,
Universit\'e de Montr\'eal, C.P. 6128, Succ.~Centre-ville,
Montr\'eal (Qu\'ebec) H3C 3J7, Canada.}
\footnotetext[3]{Department of Mathematical Sciences, University of Durham, Durham DH1 3LE, United Kingdom.}
\footnotetext[4]{email:delisle@dms.umontreal.ca}
\footnotetext[5]{email:hussin@dms.umontreal.ca}
\footnotetext[6]{email:w.j.zakrzewski@durham.ac.uk}

\date{\today}

\maketitle

\begin{abstract}
We generalize here our general procedure for constructing constant curvature   maps of 2-spheres into Grassmannian manifolds $G(m,n)$
this time concentrating our attention on maps which are non-holomorphic.
We present some expressions describing these solutions in the general case and discuss how to use these results to construct solutions of constant curvature. 
We also discuss possible values of this constant curvature.

\end{abstract}



\section{Introduction \label{intro}}
Recently, we have presented an expression 
for the Gaussian curvature of holomorphic immersions
into complex Grassmannian manifolds $G(m,n)$ \cite{newpaper}.
In this case $G(m,n)$ was described by
a $n\times m$ complex matrix field $Z$ which satisfied 
\begin{equation}
 Z^{\dagger} Z\,=\, {\mathbb I}_m, 
\label{Zprod}
\end{equation}
where ${\mathbb I}_m$ is the $m\times m$ unit matrix and as usual the symbol $\dagger$ denotes Hermitian conjugation.

We parametrized the $Z$ matrix in the following way. We introduced $\tilde Z$ a holomorphic $n\times m$ matrix obtained from a set of linearly independent holomorphic vector fields $f_1,\dots,f_m$ and $\tilde L$ a $m\times m$ matrix such that:
\begin{equation}
Z\,=\, \tilde Z \tilde L.
\label{moreh}
\end{equation}
 Such a parametrization can be called orthogonalised as it involves $Z$ that was obtained by orthogonalising the set $\{f_1,\cdots,f_m\}$. Then due to (\ref{Zprod}) 
we have defined a new matrix ${\tilde M}$ as:
\begin{equation}
{\tilde M}=({\tilde L} {\tilde L}^{\dagger})^{-1}={\tilde Z}^{\dagger}{\tilde Z}.
\label{em1}
\end{equation}
Next we have showed that  the Lagrangian density ${\cal L}$ of this holomorphic
immersion is given by
\begin{equation}
{\cal L}(Z)\,=\frac{1}{2} \,\partial_{+}\partial_{-} \,\hbox{ln}\det \tilde M,
\label{lagrangian}
\end{equation}
where the partial derivatives  ($\partial_{\pm}=\partial_{x_\pm}$) are taken with respect to  complex coordinates $x_{\pm}$. The associated  curvature of this immersion is  \cite{bolton}
\begin{equation}
{\cal K}(Z)\,= -\frac{1}{{\cal L}(Z)}\, \partial_{+}\partial_{-} \,\hbox{ln}\, {\cal L}(Z).
\label{expcurv}
\end{equation}

Thus we see that the discussion of determining admissible constant curvature holomorphic solutions of Grassmannian manifolds $G(m,n)$
 has been reduced to having to find all possible holomorphic matrices $\tilde Z$ and the corresponding curvatures that satisfy 
\begin{equation}
\det {\tilde M}\propto (1+\vert x\vert^2)^r,
\end{equation}
where the positive integer $r$ is related to the curvature by
\begin{equation}
{\cal K}=\frac{4}{r}.
\label{curvature}
\end{equation}

In our previous paper \cite{newpaper}, we have also conjectured that, for $m$ fixed, holomorphic solutions with constant 
curvature in $G(m,n)$ can be constructed for all integer values of $r$ such that 
\begin{equation}
1\leq r\leq r_{h,{\rm max}}(m,n)=m(n-m). 
\label{rhol}
\end{equation}

In this paper, we go further and look at other solutions of the Grassmannian model, the ones which are non-holomorphic. 
Thus we consider  $\tilde Z$ that is not constructed out of holomorphic vectors and for which 
the simplified formula (\ref{lagrangian}) is no longer valid. 
There are several papers in which some such solutions have already been studied.
Some early papers are explicit in the construction of these solutions \cite{early}, some more recent ones \cite{wood} are more general but less explicit.
 Our discussion, presented in this paper, provides explicit formulae for some of these solutions. In this discussion
 we  concentrate our attention on  solutions of constant curvature.
In our approach we  rely heavily on Veronese curves and we show that the admissible values of $r$ in
 the expression (\ref{curvature}) of the curvature follow an explicit rule and that they are all greater than $r_{h,max}(m,n)$, the maximal value for the holomorphic case, as given in (\ref{rhol}).

Section 2 presents a general discussion of solutions of the Grassmannian models.
In it, first we look at the simplest  model, namely $G(1,n)=\mathbb{C}P^{n-1}$, and, for completeness, we recall the general  construction of all solutions of this model. We also discuss some of their properties. These results are then used in Section 3 in which we look at solutions
of more complicated Grassmannian models. First we demonstrate which properties of the $\mathbb{C}P^{n-1}$
solutions generalise to these models and then show how our approach can be used to classify all solutions of constant curvature. Section 4 illustrates possible values of the curvature and some solutions for some Grassmannian models.

 We finish the paper with a short summary of our main
results and some conclusions.

\section{Grassmannian models}
\subsection{General discussion}

General maps of $S^2$ into a Grassmannian manifold $G(m,n), \ n>m$ are given by 
 $n\times m$ complex matrix valued fields $Z$ subject to the constraint (\ref{Zprod}). Under global $V\in U(n)$ and local $U\in U(m)$ transformations
these maps transform as
\begin{equation}
Z\,\rightarrow VZU.
\label{trans}
\end{equation}
Minimal immersions are obtained by minimizing the Lagrangian:
\begin{equation}
{\cal S}=4\int_{S^2}{\cal L}(Z)\ 
dx_+ dx_-.
\label{action}
\end{equation}
Here, $x_\pm=x\pm iy$
 are local coordinates in $\mathbb{R}^2$. The shift from $\mathbb{R}^2$ to $S^2$ will be performed by choosing a compactification of $\mathbb{R}^2$ as discussed below.
The Lagrangian density ${\cal L}$ is given by
\begin{equation}
{\cal L}(Z)\,=\,\frac{1}{2}\hbox{Tr}\left[({D_+} Z)^{\dagger}{D_+} Z+({ D_-}Z)^{\dagger}{D_-}Z\right],
\label{lagrangiancomplex}
\end{equation}
where $D_\pm$ denote covariant derivatives acting on
$Z:\mathbb{R}^2 \rightarrow G(m,n)$ and are
defined by
\begin{equation}
{D_\pm} Z =\partial_{\pm}Z-  Z Z^{\dagger} \partial_{\pm}Z.
\label{covader}
\end{equation}
The Euler-Lagrange equation corresponding to (\ref{lagrangiancomplex}) takes the form:
\begin{equation}
 {D_+} {D_-}Z+Z( {D_-}Z)^{\dagger} {D_-}Z=0.
 \label{eulerlagrangecomplex}
\end{equation}
 As we are interested in the maps of $S^2$ into the Grassmannians we have to compactify $\field{R}^2$. This we do by adding a point
at $\infty$ and this requirement `chooses' for us the boundary conditions:
\begin{equation}
\label{bc}
D_{\mu}Z\,\rightarrow\,0, \quad \hbox{as}\quad \sqrt{x^2+y^2}\,\rightarrow \,\infty
\end{equation}
sufficiently fast so that the total Lagrangian  $\mathcal{S}$ is finite. Then our maps are maps of 2-spheres into the Grassmannians 
with appropriate topological properties (see {\it i.e.} ${}^{3}$).

\subsection{Classical Solutions}

A construction of a large class of classical solutions of the Grassmannian models $G(m,n)$, which are of course minimal immersions of $S^2$, 
is well known (see {\it e.g.} \cite{wojtekbook}). This construction gives  all solutions
in the $G(1,n)$ case. For $G(m,n)$ with $m>1$, the situation is less clear but most (if not all) solutions 
can be constructed using the approach discussed in \cite{wojtekbook}. In any case, only such solutions have so far been looked at in any detail
and in this paper we restrict our attention to using them and studying their properties.

A possible way to find these solutions is to start with a set of holomorphic vectors
$f_1,\cdots, f_k$ ({\it i.e.} functions of $x_+$).
Here $k$ can be any integer up to $n-1$ (note that $n>m$).
Then one considers another set obtained from this set by taking derivatives
{\it i.e.}  $\partial_{+} f_1,\cdots, \partial_{+}  f_k$ 
and $\partial_{+} ^2f_1,\cdots, \partial_{+}^2 f_k$ and so on. Next one constructs a matrix whose columns 
are the first set, then the next one and so on. Finally, we Gram-Schmidt orthogonalise all these vectors.

Next we note that we can take any set of $m$ vectors from these orthogonalised vectors and construct from them our 
matrix $\tilde Z$. This matrix $Z$ can be shown to solve the Euler-Lagrange equation
and so defines a solution of the Grassmannian model $G(m,n)$.
If the original vectors $f_1,\,\cdots\, f_k$ are all polynomial in $x_+$ then this solution
describes an immersion of $S^2$ into $G(m,n)$.

Let us mention here a few classes of solutions derived this way:
\begin{itemize}
 \item We take $m$ holomorphic polynomial vectors ({\it i.e.} $k=m$).
       In this case we have a holomorphic solution.
       An example of such a case has been given in \cite{Macfarlane}.
\item  We start with one function ($f$) only ({\it i.e.} $k$=1).
       Then our construction will be equivalent to defining the
       operator $P_+$ as
\begin{eqnarray}
P_{+}: f \in \field{C}^N \rightarrow
P_{+}f=\partial_{+} f - \frac{f^{\dagger} \partial_{+} f }{\vert f \vert^2} f 
\label{operatorintermsoff}
\end{eqnarray}
   and then applying it up to $n-1$ times to $f$ and to the new vectors obtained 
   from it, {\it i.e.}
   \begin{equation}
    \label{repeat}
     P_+^{i} f\,=\, P_+ (P_+^{i-1} f).
   \end{equation}
A solution of the Grassmannian model $G(m,n)$ then involves taking for $Z$ any $m$ vectors
from the orthonormalized set ($\frac{f}{\vert f\vert},\,\frac{P_+f}{\vert P_+f\vert},\,\cdots\, , \frac{P_+^{n-1}f}{\vert P_+^{n-1}f\vert}$). Note that if we take the first $m$ of them the solution is holomorphic. And if we take the last $m$ of them the solution is antiholomorphic. 
But we can take any $m$ of them, say, ($\frac{f}{\vert f\vert}$, $\frac{P_+^2f}{\vert P_+^2f\vert},\cdots$).
Then the solution will be called non-holomorphic.
\end{itemize}

\medskip

In fact there are many more solutions than those described by our procedure
given above. Thus we could `miss out' some vectors from our original
set or interchange them. In such cases, there are some conditions that the vectors
have to satisfy in order that the final matrix $Z$ solves the Euler-Lagrange equation. The interested
reader can find the discussion of these conditions in the original papers and
in \cite{wojtekbook} where these papers have been referenced. Here we will restrict our attention to the cases mentioned above.

\subsection{Special case $G(1,n)=\mathbb{C}P^{n-1}$ model}

In this case the field $Z$ is a $(n\times 1)$ matrix and any solution of the Euler-Lagrange 
equation is given by
\begin{equation}
Z_i\,=\,\frac{P_+^if}{\vert P_+^if\vert},
\label{cp1}
\end{equation}
for some $i=0,\cdots, n-1$ and some holomorphic vector $f$. For the map to be from $S^2$ and not $\mathbb{R}^2$ 
the components of the vector $f$ have to be ratios of polynomials in $x_+$ \cite{wojtekbook}.  In fact,
due to the invariance (\ref{trans}) they can be given by polynomials in $x_+$.

Of these solutions those corresponding to $i=0$ are holomorphic, those corresponding to 
$i=n-1$ are anti-holomorphic and the remaining ones are 'mixed' (also called non-holomorphic in this paper).

Consider now one of these solutions, say, corresponding to a general $i$ for $i\neq0$ and $i\neq n-1$. Then
its Lagrangian density is given by \cite{wojtekbook} 
\begin{equation}
 \label{cp11}
{\cal L}(Z_i)\,=\,\frac{1}{2} \left( \frac{\vert P_+^{i+1} f\vert^2}{\vert P_+^i f\vert^2}\,+
\, \frac{\vert P_+^{i} f\vert^2}{\vert P_+^{i-1} f\vert^2}\right).
\end{equation}
Let us note that 
\begin{equation}
 \label{cp12}
\partial_{+} \partial_{-}  \ln (\vert P_+^i f\vert^2)\,=\, 
\frac{\vert P_+^{i+1}f\vert^2}{\vert P_+^i f\vert^2}\,-
\, \frac{\vert P_+^{i}f\vert^2}{\vert P_+^{i-1}f\vert^2},
\end{equation}
which, of course, is due to its topological nature.
Hence its easy to verify that
\begin{equation}
 \label{cp13}
{\cal L}(Z_i)\,=\, \frac{1}{2}\, \partial_{+}  \partial_{-}  \ln \left( \vert P_+^i f\vert^2
\vert P_+^{i-1} f\vert^4\cdots \vert P_+ f\vert^4\vert f\vert^4\right).
\end{equation}
Note that, in the holomorphic case ($i=0$), we have
\begin{equation}
 \label{cp130}
{\cal L}(Z_0)\,=\, \frac{1}{2}\, \frac{\vert P_+ f\vert^2}{\vert f\vert^2} =\frac{1}{2}\, \partial_{+}  \partial_{-}  \ln \left( \vert f\vert^2
\right).
\end{equation}

Moreover, a few lines of algebra then shows that this expression 
is much simpler if we use the formulation involving `wedge products'; namely
we note that 
\begin{equation}
 \label{cp14}
P_+^i f\,\sim \, (f\wedge\partial_{+}  f\wedge \cdots \partial_{+} ^{i-1}f)^{\dagger} (f\wedge
\partial_{+}  f\wedge \cdots \partial_{+} ^{i-1}f\wedge\partial_{+} ^i f),
\end{equation}
where $\sim$ differs from $=$ by an overall factor (up to irrelevant constants)
\begin{equation}
\frac{1}{\vert(f\wedge\partial_{+}  f\wedge  \partial_{+} ^{i-1}f)\vert^2}.
\label{cp15}
\end{equation}
To go further, we need to calculate 
\begin{equation}
\vert (f\wedge \partial_{+}  f\wedge \cdots \partial_{+} ^{i-1}f)^{\dagger} (f\wedge
 \partial_{+}  f\wedge \cdots \partial_{+} ^{i-1}f\wedge \partial_{+} ^i f)\vert^2.
\label{a}
\end{equation}
This, as it is easy to check, is the product of determinants $M_{i+1}$ and $M_{i}$,
where
\begin{equation}
M_i\,=\,
\det \left(\begin{array}{cccc}
\vert f\vert^2& f^{\dagger}\partial_{+}  f&...& f^{\dagger}\partial_{+} ^{i-1} f\\
(\partial_{+}  f)^{\dagger}f& \vert \partial_{+}  f\vert^2&...&(\partial_{+}  f)^{\dagger}\partial_{+} ^{i-1} f\\
\vdots& \vdots& \vdots&\vdots\\
(\partial_{+} ^{i-1}  f)^{\dagger}f& (\partial_{+} ^{i-1}  f)^{\dagger}\partial_{+} f&...&\vert \partial_{+} ^{i-1}  f\vert^2
\end{array}\right)=\prod_{k=0}^{i-1}\vert P_+^kf\vert^2, \quad M_0\,=\,1. 
\label{det}
\end{equation}

However, this is exactly what we need for rewriting $\cal L$ in a simple way.
Using the expressions above it becomes
\begin{equation}
 \label{cp16}
 {\cal L}(Z_i)\,=\,\frac{1}{2}\, \partial_{+} \partial_{-} \ln\,( M_{i+1} M_{i} ).
\end{equation}
Thus we see that taking $i=0$, we retrieve the holomorphic case (\ref{cp130}).

Next consider constant curvature solutions. This implies that we require to have 
\begin{equation}
M_{i+1} M_{i}\propto (1+\vert x\vert^2)^{r_{i}(1,n)}
\end{equation}
in which case the corresponding curvature $\mathcal{K}$ is given by ${\cal K}(Z_i)=\frac{4}{r_{i}(1,n)}$. In the following, we  determine all possible values of ${r_{i}(1,n)}$, where the label $i$ is related to the label of the solutions $Z_i$ and $(1,n)$ refers to the $G(1,n)$ model.
We already know that the only holomorphic solutions with constant curvature in $G(1,n)=\mathbb{C}P^{n-1}$ are the Veronese minimal spheres \cite{bolton, wojtekbook}. 

Let us next show that, using the projector formalism, it is easy to determine all the possible values of the corresponding curvatures for non-holomorphic solutions.

Starting from the holomorphic Veronese curve $f^{(n)}: S^2 \to \field{C}P^{n-1}$ :
\begin{equation}
{f^{(n)}}
=\left(1, \sqrt{\left(\begin{array}{c}
n-1 \\
1 \end{array}\right)}\, x_+\,, \dots, 
\sqrt{\left(\begin{array}{c}
n-1 \\
r \end{array}\right)}\, {x_+}^r\,, \dots,
{x_+}^{n-1}
\right)^T\, ,
\label{veroneseseq}
\end{equation}
with constant curvature ${\cal K}(Z_i)=\frac{4}{n-1}$,
we get a set of  linearly independent solutions given by  $\{ \frac{{P}_+^i f^{(n)}}{\vert {P}_+^i f^{(n)} \vert}, i=0,\dots, n-1\}$ following the procedure described above.  To this set corresponds a set of orthogonal projectors 
$P_{i}(f^{(n)})$ defined as follows:
\begin{equation}
P_{i}(f^{(n)}):=Z_i Z_i^{\dagger}=\frac{{P}_+^i f^{(n)}  {({P}_+^i f^{(n)}})^{\dagger}}
{\vert {P}_+^i f^{(n)} \vert^2},\,\,\, 
i =0,1,...,n-1\,.
\label{concproject}
\end{equation}
We have shown in \cite{HYZ} that, for each solution in this set, the  curvature $\mathcal{K}(Z_i)$ is related to the following quantity 
 \begin{equation}
r_{i}(1,n)=A(n, P_{i}(f^{(n)}))=(n-1)+2i (n-1-i).
\label{rnk}
\end{equation}
This formula may be easily recovered from the expression of ${\cal L}(Z_i)$ given in (\ref{cp13}) using the fact that 
\begin{equation}
\vert P_{+}^i(f^{(n)})\vert^2=\frac{(n-1)!\,i!}{(n-1-i)!}(1+\vert x\vert^2)^{n-1-2i}.
\label{pplusf}
\end{equation}
We see that $r_{0}(1,n)=r_{n-1}(1,n)=n-1$. This is the minimal value of $r_{i}(1,n)$ in the set and  it corresponds to the largest constant curvature ${\cal K}=\frac{4}{n-1}$ for the holomorphic and anti-holomorphic solutions. We also have
\begin{equation}
A(n,P_{n-1-i}(f^{(n)}))=A(n,P_{i}(f^{(n)})),
\label{rnk}
\end{equation}
which implies that among the  projectors of the set only `a half of them' give rise to different curvatures. More precisely, for $n=2p$ we consider only the projectors $P_i(f^{(n)})$ with $i=1,\dots,p-1$, while for $n=2p+1$, we take $i=1,\dots,p$ for non equivalent non-holomorphic solutions.

 Let us note that for non equivalent solutions, we have 
\begin{equation}
r_{i+1}(1,2p)-r_{i}(1,2p)\geq 4, \ i=0,\dots, p-2,
\end{equation}
\begin{equation}
r_{i+1}(1,2p+1)-r_{i}(1,2p+1)\geq 2, \ i=0,\dots, p-1,
\end{equation}
which was deduced from
\begin{equation}
 r_{i+1}(1,n)-r_i(1,n)=2(n-2i-2).
\end{equation}

We thus get a higher bound for the values of $r$ which is obtained as
\begin{equation}
r_{p-1}(1,2p)=2p^2-1,\quad r_{p}(1,2p+1)=2p(p+1).
\end{equation}

Finally, it is easy to show that, for $p,q\in\mathbb{N}$ and $q<p$, 
\begin{eqnarray}
r_{p-q}(1,2p)&=&r_{0}(1,2(p^2-q(q-1))),\\
r_{p-q}(1,2p+1)&=&r_{0}(1, 2(p(p+1)-q^2)+1).
\end{eqnarray}
This result relates non-holomorphic solutions of constant curvature to a holomorphic solution of a higher dimensional Grassmannian $G(1,N)$ (for $N>n$).

Let us illustrate these results by some explicit examples:

\begin{itemize}
 \item For $\field{C}P^{2} (n=3)$, we get only one non-holomorphic solution corresponding to $P_{1}(f^{(3)})$ for which $r_1(1,3)=4$ (the same value as for $P_{0}(f^{(5)})$). Let us recall that holomorphic solutions are obtained  for $r=1, 2$.
\item For $\field{C}P^{3} (n=4)$, we get one non-holomorphic solution corresponding to $P_{1}(f^{(4)})$ for which $r_1(1,4)=7$ (the same value as for $P_{0}(f^{(8)})$ and also, by embedding,  the preceding one {\it i.e.} of the  $\field{C}P^{2}$ field  with $r=4$). Note that holomorphic solutions are easily found for $r=1, 2, 3$.
\item For $\field{C}P^{4} (n=5)$, we have two new non-holomorphic solutions with $r_1(1,5)=10$ and $r_2(1,5)=12$ and the embeddings with $r=4,7$. Of course the holomorphic solutions are obtained  for $r=1, 2, 3, 4$.
\item For $\field{C}P^{5} (n=6)$, we have two new non-holomorphic solutions with $r_1(1,6)=13$ and $r_2(1,6)=17$ and the embeddings with $r=4, 7, 10, 12$. The holomorphic solutions exist for $r=1, 2, 3, 4, 5$.
\end{itemize}

\section{Non-holomorphic solutions for $G(m,n)$}

As we have said above there are many non-holomorphic solutions of $G(m,n)$ models.
In section 2, we discussed a class of them obtained by applying the $P_+$
operator defined  in (\ref{operatorintermsoff}). Then if we apply it to a holomorphic vector $f_1$
several times - we obtain vectors $P_+^if_1$.
In what follows we shall take $f_1=f^{(n)}$ as the Veronese curve (\ref{veroneseseq}). Any pair of these normalized vectors
gives a solution of $G(2,n)$, any triple  produces a solution of $G(3,n)$ {\it etc}.

A generic solution $Z$ of $G(m,n)$, made of $m$ normalized independent vectors  taken among the set $\{f^{(n)}, P_+f^{(n)},\dots, P_+^{n-1}f^{(n)}\}$, gives rise to a projector  $P=ZZ^{\dagger}=\sum_{i=0}^{n-1}\alpha_iP_i$, where the constants $\alpha_i$ take  values $0$ or $1$ and $P_i$ acts on $f^{(n)}$ as in (\ref{concproject}).  We have shown in \cite{HYZ}  that, for such a generic solution, the Lagrangian density is given by
\begin{equation}
\mathcal{L}(Z)=\frac{1}{2}\sum_{i=1}^{n-1}(\alpha_{i-1}-\alpha_i)^2\frac{\vert P_+^{i}f^{(n)}\vert^2}{\vert P_+^{i-1}f^{(n)}\vert^2}.
\end{equation} 
This expression  can  be rewritten in the following more compact form, using the expression $M_i$  given in (\ref{det}), as
\begin{equation}
\mathcal{L}(Z)=\,\frac{1}{2} \partial_+\partial_-\ln \prod_{i=1}^{n-1}M_i^{(\alpha_{i-1}-\alpha_i)^2}.\label{genLagrangian}
\end{equation}
In particular, if  $\alpha_0=\dots=\alpha_{m-1}=1$ and $\alpha_m=\dots=\alpha_{n-1}=0$, we get easily
\begin{equation}
\mathcal{L}(Z)=\frac{1}{2}\partial_+\partial_-\ln M_m,
\label{holomn}
\end{equation}
which corresponds to a holomorphic solution of the $G(m,n)$ model as discussed in \cite{newpaper}.

In order to get constant curvature solutions, we  require that
\begin{equation}
 \prod_{i=1}^{n-1}M_i^{(\alpha_{i-1}-\alpha_i)^2}\propto (1+\vert x\vert^2)^r.
\end{equation}
But the $M_i$'s are products of consecutives $\vert P_+^kf^{(n)}\vert^2$, which implies that each $P_+^kf^{(n)}$ must be such that
\begin{equation}
\vert P_+^kf^{(n)}\vert^2\propto(1+\vert x\vert^2)^{r_k}.
\end{equation}

For the Veronese sequence with $f^{(n)}$ given by (\ref{veroneseseq}), using (\ref{pplusf}), we get
\begin{equation}
\partial_+\partial_-\ln M_i=\frac{i(n-i)}{(1+\vert x\vert^2)^2}\label{calc}
\end{equation}
and so (\ref{genLagrangian}) may be rewritten as
\begin{equation}
\mathcal{L}(Z)=\frac{1}{2}\partial_+\partial_-\ln(1+\vert x\vert^2)^{r(m,n)},\quad r(m,n)=\sum_{i=1}^{n-1}i(n-i)(\alpha_{i-1}-\alpha_i)^2.
\end{equation}
Hence, we get a constant curvature solution of our Grassmannian $G(m,n)$ model with
\begin{equation}
\mathcal{K}(Z)=\frac{4}{r(m,n)}.
\label{constKmn}
\end{equation}
Note that for the holomorphic solution corresponding to (\ref{holomn}), we get $r_{h, {\rm max}}(m,n)=m(n-m)$ as expected (see (\ref{rhol})).

 Let us mention that due to the property $G(m,n)\simeq G(n-m,n)$ (which is easy to see in the projector formulation), we will consider solutions only for the model $G(m,n)$ (with $m=1,2,...[\frac{n}{m}]$) but our construction will also give the solutions for the models with larger $m$. Thus, in particular, we have already all the solutions of $G(n-1,n)\simeq \mathbb{C}P^{n-1}$.

In order to make our discussion more intuitive, let us first discuss in detail the $G(2,n)$ model. We, thus, have to distinguish the cases when
the two $P_+^lf^{(n)}$ vectors forming the solution have their corresponding $l$'s differing by 1 or not.
The reason for this is simple: the Lagrangian density of the vectors which differ 
by more than one is purely additive; it is simply a sum of Lagrangian densities
of the corresponding $\mathbb{C}P^{n-1}$. We will then show how this gets modified for larger values of $m$.

\subsection{$G(2,n)$}

We take a solution of $G(2,n)$ of the form
\begin{equation}
 Z_{i,j}^{(n)}=\left(
      \frac{P_+^i f^{(n)}}{\vert P_+^i f^{(n)}\vert},  \frac{P_+^j f^{(n)}}{\vert P_+^j f^{(n)}\vert}\right),
       \end{equation}
       where $f^{(n)}$ is the Veronese curve (\ref{veroneseseq}) for any integer $i, j$ such that $i\neq j$ and $0\leq i\leq n-2, \ 1\leq j\leq n-1$.
       
It is easy to see that the case $j>i+1$ leads to:
\begin{equation}
{\cal L}( Z_{i,j}^{(n)})\,={\cal L}( Z_i)+{\cal L}(Z_j)\,=\, \frac{1}{2} \partial_{+} \partial_{-} \ln\,( M_{i+1}  M_{i}M_{j+1} M_{j}),
\label{elij}
\end{equation}
where $Z_i$ is defined in (\ref{cp1}), and the constant curvature is given as in (\ref{constKmn}) with
\begin{equation}
\label{highr}
r(2,n)=r_{i,j}(2,n)=r_{i}(1,n)+r_{j}(1,n)=2\left(n-1+i(n-1-i)+j(n-1-j)\right).
\label{rij}
\end{equation}

Next we consider the case of consecutive projectors, \textit{i.e.} when $j=i+1$. The calculation of the Lagrangian density gives
\begin{equation}
 \label{Glplusone}
{\cal L}( Z_{i,i+1}^{(n)})\,=
\,\frac{1}{2} \left(\frac{\vert P_+^{i}f\vert^2}{\vert P_+^{i-1}f\vert^2}\,+\,
\frac{\vert P_+^{i+2}f\vert^2}{\vert P_+^{i+1}f\vert^2}\right)\,=\,\frac{1}{2} \partial_{+} \partial_{-} \ln\,( M_{i+2}  M_{i})
\label{eliiplus1}
\end{equation}
and the constant curvature is given as in (\ref{constKmn}) with
\begin{equation}
r(2,n)=r_{i}(2,n)=2(n-2+ i(n-2-i)),
\label{riiplus1}
\end{equation}
for $0\leq i\leq n-2$.
The holomorphic case ($i=0$) is included in formula (\ref{Glplusone}) and reduces to expression (\ref{holomn}) with $m=2$.

Due to the way the set of solutions was constructed, we have a relation \cite{HYZ} between $P_+^{i}f$ and $P_+^{n-1-i}f$ through complex conjugation.  We thus have equivalent solutions $Z_{i,i+1}^{(n)}\sim Z_{n-2-i,n-1-i}^{(n)}$ and  $Z_{i,j}^{(n)}\sim Z_{n-1-j,n-1-i}^{(n)}$ . This leads to
\begin{equation}
r_i(2,n)=r_{n-2-i}(2,n), \quad  r_{i,j}(2,n)=r_{n-1-j,n-1-i}(2,n).
 \label{symmetry}
\end{equation}

Let us now look at some properties of these expressions for different values of the parameter $r$. 
 First, due to the relation (\ref{symmetry}), non equivalent solutions of the type $Z_{i,i+1}^{(n)}$ are obtained for $i=0,1,\dots, p-1$ where $p=[\frac{n}{2}]$ and $p \geq 2$. We thus have,  for $i=0,1,\dots, p-1$, 
\begin{equation}
 r_{i+1}(2,2p)-r_i(2,2p)\geq2,\quad r_{i+1}(2,2p+1)-r_i(2,2p+1)\geq4.
\end{equation}
Indeed, this result follows from
\begin{equation}
 r_{i+1}(2,n)-r_i(2,n)=2(n-2i-3)\geq2(n-2p+1).
\end{equation} 

Second, in the set $ \{r_{i,j}(2,n)\}$ that leads to non equivalent solutions, we see that the minimal value of $r$ for non holomorphic solutions of $G(2,n)$, for $n>4$, is
\begin{equation}
r_{0,n-1}(2,n)=2(n-1)> r_0(2,n)=2(n-2).
\end{equation}
Moreover, for $i=1,\cdots, p-1$ with $p=[\frac{n}{2}]$ and $p\geq 2$, we have
\begin{equation}
 r_i(2,n)-r_{0,n-1}(2,n)\geq2(n-4).
\end{equation}

Third, we can show that we get distinct non-holomorphic solutions with the same value of $r$. Indeed, from the definition of the $r_{i,j}(2,n)$, we have, in particular, that
\begin{equation}
r_{i,j}(2,n)=r_{i,k}(2,n),
\end{equation}
for $k=n-1-j$.
Since $j$ is at least equal to $2$ and $k>j$, we get $n>5$. For example, for $n=6$, we have $r_{0,2}(2,6)=r_{0,3}(2,6)=22$. For $n=7$, we have $r_{0,2}(2,7)=r_{0,4}(2,7)=28$.
Moreover, we can have $r_i(2,n)=r_{j,k}(2,n)$ for $n>6$ and some values of $i,j,k$. For example, for $n=7$, we have $r_{0,5}(2,7)=r_2(2,7)=22$. Furthermore, as $n$ increases identical values of $r$ appear for a larger number of distinct solutions.

Finally, we give the higher bound for the values of $r$ for non-holomorphic solutions of $G(2,n)$. 
We have
\begin{equation}
r_{p-2,p}(2,2p)=2(2p^2-3),\quad r_{p-1,p+1}(2,2p+1)=4(p^2+p-1).
\label{higher}
\end{equation}
Thus the lowest value of the curvature is given by the appropriately chosen two projectors with $i$ and $j$ differing by 2.

Let us add that due to (\ref{symmetry}), we see that values of $r_{i,j}(2,n)$ for non equivalent solutions are given by
\begin{equation}
r_{i,j}(2,2p), \  i= 0, \dots, p - 2, \quad j= i + 2, \dots, 2 p - 1 - i,
\end{equation}
\begin{equation}
r_{i,j}(2,2p+1), \  i= 0, \dots, p - 1, \quad j= i + 2, \dots, 2 p  - i.
\end{equation}

To prove (\ref{higher}), we note from (\ref{rij}) that the maximal value of $r$ is given by the values of $i$ and $j$ which maximise $i(n-1-i)+j(n-1-j)$. It is clear that the value of $i$ which maximises $i(a-i)$ is given by $i=\frac{a}{2}$.
As $i$ has to be an integer, this value is reached when $n$ is odd since $a=n-1$ is then even.
For $n$ even, the maximum value is given by the nearest integer {\it i.e.} for $i=\frac{a\pm 1}{2}$.
As $i$ and $j$ have to satisfy the condition $j>i+1$, we have to take $j=i+2$ and place $i$ and $j$ as close to $\frac{n-1}{2}$ as possible.
This, as can be easily checked, gives the values mentioned above.

We also have to prove that the values in (\ref{higher}) are higher then the largest value of the $r_i(2,n)$.  Indeed, the largest value of $r$, when $n$ is even, corresponds to $i=\frac{n-2}{2}$ and then $r_{max}=\frac{n^2-4}{2}$.
For $n$ odd the corresponding value is given by $r_{max}=\frac{n^2-5}{2}$ where $i=\frac{n-3}{2}$. 

At this stage, we could illustrate these results for some values of $n$. 

\begin{itemize}
 \item For $n=3$, the duality property leads to $G(2,3)\simeq \mathbb{C}P^2$.
 
\item For $n=2p=4$, in the case of consecutive projectors, we have only two non equivalent solutions $Z_{0,1}^{(4)}, \ Z_{1,2}^{(4)}$ where $Z_{0,1}^{(4)}$ is holomorphic with $r_0(2,4)=4$ and $Z_{1,2}^{(4)}$ is non holomorphic with $r_1(2,4)=6$. For non consecutive projectors, the completion relation $\sum_{i=0}^3P_i=\mathbb{I}$, leads to $Z_{0,3}^{(4)}\sim Z_{1,2}^{(4)}$ and $Z_{1,3}^{(4)}\sim Z_{0,2}^{(4)}$. Thus the only remaining case is $Z_{0,2}^{(4)}$ with $r_{0,2}(2,4)=10$.

The solutions $Z_{1,2}^{(4)}$ and $Z_{0,2}^{(4)}$ correspond to the two last solutions presented in theorem B in \cite{zhen}. The missing solution corresponding to $r=2$ is actually the solution corresponding to the direct sum of $\mathbb{C}P^1\oplus\mathbb{C}P^1$.

\item For $n=2p+1=5$, in the case of consecutive projectors, the symmetry property (\ref{symmetry}) leads to
two non equivalent solutions $Z_{0,1}^{(5)}, \ Z_{1,2}^{(5)}$ where $Z_{0,1}^{(5)}$ is holomorphic with $r_0(2,5)=6$ and $Z_{1,2}^{(5)}$ is non holomorphic with $r_1(2,5)=10$. For non consecutive projectors, the symmetry property leads to $Z_{0,2}^{(5)}\sim Z_{2,4}^{(5)}$ and $Z_{0,3}^{(5)}\sim Z_{1,4}^{(5)}$. The non equivalent solutions are $Z_{0,2}^{(5)}, \ Z_{0,3}^{(5)},\ Z_{0,4}^{(5)}, \ Z_{1,3}^{(5)}$ with
$r_{0,2}(2,5)=16$, $r_{0,3}(2,5)=14$, $r_{0,4}(2,5)=8$ and $r_{1,3}(2,5)=20$.

Using $G(2,5)\simeq G(3,5)$, the above example gives all the possible values of $r$ for the $G(3,5)$ model.

\end{itemize}




\subsection{$G(m,n)$ for $m>2$ }
The $G(3,n)$ model is strongly related to the $G(2,n)$ and $G(1,n)$ cases. Indeed, we have to distinguish three cases: three isolated projectors, two consecutive projectors and one isolated projector and three consecutive projectors. Explicitly, we have ($i<j<k$)
\begin{equation}
 Z_{i,j,k}^{(n)}=\left(
      \frac{P_+^i f^{(n)}}{\vert P_+^i f^{(n)}\vert},  \frac{P_+^j f^{(n)}}{\vert P_+^j f^{(n)}\vert}, \frac{P_+^k f^{(n)}}{\vert P_+^k f^{(n)} \vert}\right),
       \end{equation}
for $(i,j,k)$ with $j>i+1$ and $k>j+1$ (the first case), $(i,j,k)=(i,j,j+1)$ with $j>i+1$ and $(i,j,k)=(i,i+1,k)$ with $k>i+2$ (the second case) and, finally, $(i,j,k)=(i,i+1,i+2)$ (the third case), for $0\leq i\leq n-1$.

In the first case, we obtain the Lagrangian density
\begin{equation}
\mathcal{L}( Z_{i,j,k}^{(n)})=\frac{1}{2}\partial_+\partial_-\ln( M_{i+1}M_i M_{j+1}M_j M_{k+1}M_k),
\end{equation}
with corresponding $r$ given by
\begin{equation}
r_{ijk}(3,n)=3(n-1)+2i(n-1-i)+2j(n-1-j)+2k(n-1-k).
\end{equation}
For the second case, we have two possibilities:
\begin{equation}
\mathcal{L}( Z_{i,j,j+1}^{(n)})=\frac{1}{2}\partial_+\partial_-\ln (M_{i+1}M_i M_{j+2}M_j),
\end{equation}
and $r$ given by
\begin{equation}
r_{i,i+1,k}(3,n)=3n-5+2i(n-1-i)+2j(n-2-j).
\end{equation}
and
\begin{equation}
\mathcal{L}( Z_{i,i+1,k}^{(n)})=\frac{1}{2}\partial_+\partial_-\ln (M_{i+2}M_i M_{k+1}M_k),
\end{equation}
and $r$ given by
\begin{equation}
r_{i,i+1,k}(3,n)=3n-5+2i(n-2-i)+2k(n-1-k).
\end{equation}

For the third case,
\begin{equation}
\mathcal{L}( Z_{i,i+1,i+2}^{(n)})=\frac{1}{2}\partial_+\partial_-\ln( M_{i+3}M_i),
\end{equation}
with
\begin{equation}
r_{i,i+1,i+2}(3,n)=3(n-3)+2i(n-3-i).
\end{equation}

Again, we can obtain upper bounds on the value of $r$. Explicitly, we get for $n\geq5$
\begin{equation}
r_{p-3,p-1,p+1}(3,2p)= 6 p^2-19,\quad r_{p-2,p,p+2}(3,2p+1)= 6 p^2 + 6 p -16.
\end{equation}

This can be easily generalized to the $G(m,n)$ model given the Lagrangian densities of the different solutions for $G(k,n)$ for $k<m$. Only one case is missing, the case were we have the sum of $m$ consecutive projectors. In this case, the Lagrangian density is given by
\begin{equation}
\mathcal{L}(Z_{i,i+1,\cdots ,i+m-1}^{(n)})=\frac{1}{2}\partial_+\partial_-\ln (M_{i+m}M_i),
\end{equation}
with $r$ given by, using formula (\ref{calc}),
\begin{equation}
r_{i,i+1,\cdots ,i+m-1}(m,n)=m(n-m)+2i(n-m-i).
\end{equation}
We see that for $i=0$, we retrieve the values of  $r$ corresponding to the holomorphic solution namely $r_{0,\cdots ,m-1}=m(n-m)$. 

In the $G(m,n)$ model with $n\geq2m-1$, the upper bound of the different values of $r$ is given by:
\begin{eqnarray}
r_{p-m,p-m+2,\cdots,p+m-2}(m,2p)=\frac{1}{3}m(6p^2-2m^2-1)\label{uppermn},\\ r_{p-m+1,p-m+3,\cdots,p+m-1}(m,2p+1)=\frac{2}{3}m(1-m^2+3p(1+p)).
\end{eqnarray}
The proof is similar to the one used in the special case of $m=2$. Note that the condition $n\geq2m-1$ is crucial in our analysis. Indeed, in the case $n=2p$, it is equivalent to $p-m\geq0$, which ensures the existence of the projector
\begin{equation}
 P_{p-m}+P_{p-m+2}+\cdots+P_{p+m-2}.
\end{equation}
We thus see that equation (\ref{uppermn}) gives the upper bounds for the $G(m,2p)$ models only for the values $1\leq m\leq p$, but using the duality property $G(m,2p)\simeq G(2p-m,2p)$ we get all of them. The reasoning is similar for the odd case $n=2p+1$.

We finish this section with the following comments: Given that $G(m,n)\simeq G(n-m,n)$, we see that, in order to get new results and solutions which are not related to the lower dimensional Grassmannians $G(i,n)$ with $i\leq m-1$, we have to impose $n-m\geq m$ or $n\geq 2m$. This means that, for $m$ fixed, the minimal value of $n$ is given by $n=2m$.

Moreover, in the case of  $G(m,2m)$, we get a set of  $\frac{1}{2}{2m\choose m}-1$ non-holomorphic and non-equivalent solutions.  Indeed, we can construct a total of ${2m\choose m}$ projectors in $G(m,2m)$. Using the completion relation $\sum_{k=1}^mP_{i_k}=\mathbb{I}-\sum_{k=m+1}^{2m}P_{i_k}$, we get $\frac{1}{2}{2m\choose m}$ non-equivalent solutions. We get the desired result by removing the holomorphic solution.

\section{Constant curvatures for some Grassmannian models}
In this section, we summarize the possible values of the constant $r(2,n)$ appearing in the curvature $\mathcal{K}=\frac{4}{r(2,n)}$ for the $G(2,n)$ model with $n=4,5,6,7$. 
The possible values of $r(2,n)$ are listed in the following table. 
\begin{center}
 \begin{tabular}{|c|c|c|c|c|}
  \hline
$G(2,n)$&$G(2,4)$&$G(2,5)$&$G(2,6)$&$G(2,7)$\\
\hline
$r_0$&4&6&8&10\\
\hline
$r_1$&6&10&14&18\\
\hline
$r_2$&--&--&16&22\\
\hline
$r_{02}$&10&16&22&28\\
\hline
$r_{03}$&--&14&22&30\\
\hline
$r_{04}$&--&8&18&28\\
\hline
$r_{05}$&--&--&10&22\\
\hline
$r_{06}$&--&--&--&12\\
\hline
$r_{13}$&--&20&30&40\\
\hline
$r_{14}$&--&--&26&38\\
\hline
$r_{15}$&--&--&--&32\\
\hline
$r_{24}$&--&--&--&44\\
\hline
 \end{tabular}
\end{center}

This puts together all the values mentioned in sections 2 and 3. As mentioned above, in the $G(2,6)$ model, we see that we obtain two non equivalent non-holomorphic solutions $Z_{02}^{(6)}$ and $Z_{03}^{(6)}$ of the same curvature $\mathcal{K}=\frac{2}{11}
$. Here are the explicit expressions of these solutions
\begin{eqnarray*}
 Z_{02}^{(6)}=\frac{1}{(1+\vert x\vert^2)^{\frac{5}{2}}}\left(\begin{array}{cc}
1&\sqrt{10}x_-^2\\
\sqrt{5}x_+&\sqrt{2}x_-(-2+3\vert x\vert^2)\\
\sqrt{10}x_+^2&1+3\vert x\vert^4-6\vert x\vert^2\\
\sqrt{10}x_+^3&x_+(3+\vert x\vert^4-6\vert x\vert^2)\\
\sqrt{5}x_+^4&\sqrt{2}x_+^2(3-2\vert x\vert^2)\\
x_+^5&\sqrt{10}x_+^3
\end{array}\right), \\ 
Z_{03}^{(6)}=\frac{1}{(1+\vert x\vert^2)^{\frac{5}{2}}}\left(\begin{array}{cc}
1&-\sqrt{10}x_-^3\\
\sqrt{5}x_+&\sqrt{2}x_-^2(3-2\vert x\vert^2)\\
\sqrt{10}x_+^2&-x_-(3+\vert x\vert^4-6\vert x\vert^2)\\
\sqrt{10}x_+^3&1+3\vert x\vert^4-6\vert x\vert^2\\
\sqrt{5}x_+^4&\sqrt{2}x_+(2-3\vert x\vert^2)\\
x_+^5&\sqrt{10}x_+^2
\end{array}\right).
\end{eqnarray*}

 A similar exercise can be done for the $G(3,n)$ model for $n=6,7$. Indeed, for $G(3,6)$ we get
\begin{eqnarray*}
 \{r_{012}\}=\{9\},\quad \{r_{013},r_{014},r_{015}\}=\{25,21,13\},\\
\{r_{023},r_{034},r_{045}\}=\{21,19,13\},\quad \{r_{024},r_{025},r_{035}\}=\{35,27,27\}
\end{eqnarray*}
and for the $G(3,7)$ model, we have
\begin{eqnarray*}
 \{r_{012},r_{123},r_{234}\}=\{12,18,20\},\\ \{r_{013},r_{014},r_{015},r_{016},r_{124},r_{125}\}=\{34,32,26,16,40,34\},\\
\{r_{023},r_{034},r_{045},r_{134}\}=\{28,28,24,38\},\\ \{r_{024},r_{025},r_{026},r_{035},r_{036},r_{135}\}=\{50,44,34,46,36,56\}.
\end{eqnarray*}

\section{Further Comments and Conclusions}

In this paper we have generalised the results of \cite{newpaper} to non-holomorphic immersions of $S^2$ into Grassmannians. Some of our results coincide with the results obtained some time
ago (see the references in \cite{wojtekbook}) but at that time the emphasis was on different aspects of this problem.
Some of our results are, however,  more general and more explicit.  Given the mathematical 
interest in $S^2$ immersions into Grassmannians \cite{zhen} we thought it is worthwhile to look at these `older' expressions and rederive them in a new setting.
Moreover, our procedure is simpler and, in a way, more explicit. 
In particular, it can be used to check with ease whether a given
immersion has a constant curvature or not (see our work in \cite{newpaper}).

In addition, it also shows very clearly how to go further and generalize it to the study of immersions into more general (larger) Grassmannians.  
This problem is currently under investigation.

Let us finish by mentioning that in this work we can also exploit the following observation.  Consider, for example, 
the solutions of the $G(2,n)$ model and note that we can obtain some of them by the following simple procedure: given two vector fields $f\in\mathbb{C}P^{k-1}$ and $g\in\mathbb{C}P^{l-1}$ such that $k+l=n$, one
can construct a solution of $G(2,n)$ by taking
\begin{equation}
Z_{ij}=\left(\begin{array}{cc}
\frac{P_+^if}{\vert P_+^if\vert}&0\\
 0&\frac{P_+^jg}{\vert P_+^jg\vert}\end{array}\right),
\end{equation}
where $0\leq i\leq k-1$ and $0\leq j\leq l-1$. The lagrangian density, as given in (\ref{lagrangiancomplex}), corresponding  to $Z_{ij}$ can be easily calculated and we get
\begin{equation}
 \mathcal{L}(Z_{ij})=\mathcal{L}(Z_i^f)+\mathcal{L}(Z_j^g),
\end{equation}
where $Z_i^f=\frac{P_+^if}{\vert P_+^if\vert}$ and $Z_j^g=\frac{P_+^jg}{\vert P_+^jg\vert}$. We thus see that if $f$ and $g$ are the Veronese sequences in $\mathbb{C}P^{k-1}$ and $\mathbb{C}P^{l-1}$, respectively, then we get						
\begin{equation}
 \mathcal{L}(Z_{ij})=\frac{r_i(1,k)+r_j(1,l)}{(1+\vert x\vert^2)^2},
\end{equation}
with corresponding constant curvature $\mathcal{K}=\frac{4}{r_i(1,k)+r_j(1,l)}$.

\section*{Acknowledgments}
This work has been supported in part by  research grants
from NSERC of Canada. LD also acknowledges a FQRNT fellowship.


\end{document}